  \documentclass[multi]{cambridge6Atight}
  \usepackage{natbib}

  \usepackage{floatpag}
  \rotfloatpagestyle{empty}

  \usepackage{amsthm}
  \usepackage{graphicx}
  \usepackage{mathptmx}

  \usepackage{makeidx}
  \makeindex

 \copyrightline{Reprinted from \textit{The Impact of Binaries on Stellar Evolution}, Beccari G. \& Boffin H.M.J. (Eds.), \copyright\ 2019 Cambridge
    University Press.}

\begin{document}


  \alphafootnotes
   \author[Ulisse Munari]
    {\sc Ulisse Munari}
  \chapter{The Symbiotic Stars}

  \arabicfootnotes

  \contributor{Ulisse Munari
    \affiliation{INAF National Institute of Astrophysics, Astronomical Observatory of Padova, 36012 Asiago (VI), Italy}}

\section{Symbiotic Stars: Binaries accreting from a Red Giant}
\label{MunariSec1}

When \citet{Merrill} discovered CI Cyg\index{CI Cyg} and AX Per\index{AX
Per}, the first known {\it symbiotic stars} (hereafter
SySts)\index{symbiotic star}, they were puzzled (in line with the wisdom of
the time, not easily contemplating stellar binarity) by the co-existence in
the 'same' star of features belonging to distant corners of the HR diagram:
the TiO bands typical of the coolest M giants, the HeII 4686 \AA\ seen only
in the hottest O-type stars, and an emission line spectrum matching that of
planetary nebulae\index{planetary nebula} (hereafter PN).  All these
features stands out prominently in the spectrum of CI Cyg shown in Figure~6.1
together with its light-curve displaying a large assortment of different
types of variability, with the spectral appearance changing in pace (a
brighter state usually comes with bluer colors and a lower ionization).

  \begin{figure}
    \includegraphics[scale=0.7]{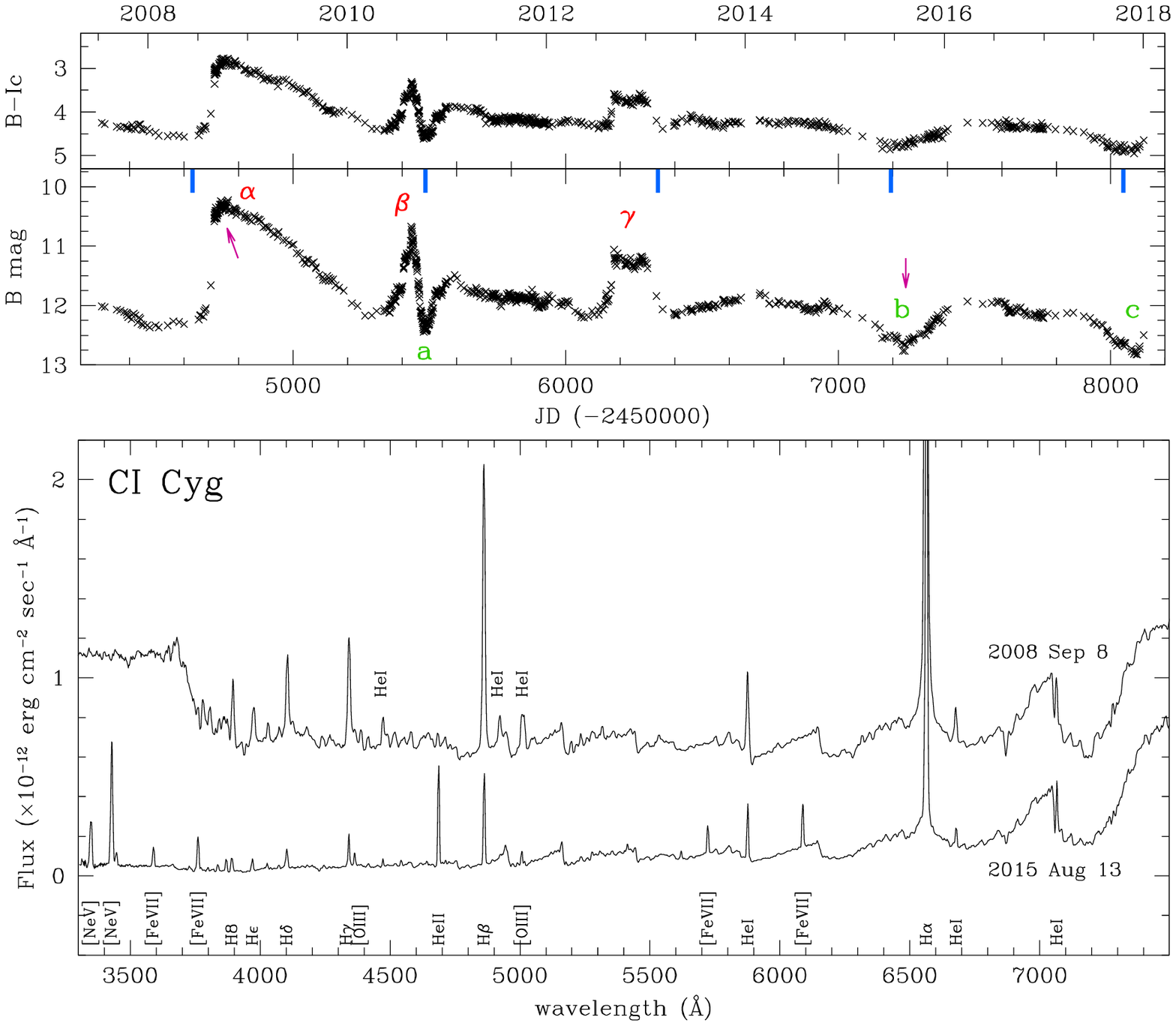}
    \caption[Photometric and spectroscopic behavior of CI Cyg]
      {Photometric and spectroscopic behavior of the prototype symbiotic
       star CI Cyg.  \textit{Upper panels:} the arrows mark the time for the
       two spectra shown, and the thick bars the passage at superior
       conjunction of the WD in the 855 days orbit around the M giant
       companion.  The letter '$a$' indicates an eclipse of the outbursting
       WD, while '$b$' and '$c$' minima in the irradiation-modulated
       light-curve.  $\alpha$, $\beta$ and $\gamma$ mark three separate
       outbursts of different amplitude, shape and duration.  The fact that
       the $B$$-$$I$ color evolution follows the $B$ light-curve indicates
       how the M giant keeps stable while the activity resides entirely with
       the WD and the circumstellar gas ionized by it.  \textit{Bottom
       panel:} spectrum at outburst peak (2008 Sep 8) and at irradiation
       minimum (2015 Aug 13).  Note the huge diversity in ionization degree
       (HeII, $[$NeV$]$, $[$FeVII$]$), integrated flux of emission lines,
       and emission in the Balmer continuum (for this and all the following
       figures spectroscopy is provided by Asiago and Varese telescopes, and
       photometry by ANS Collaboration).}
    \label{Munari1Fig:Fig1}
     \end{figure}
     
A great incentive to the study of SySts was provided in the 1980ies by the
first conference \citep{Friedjung} and monograph \citep{Kenyon86} devoted
entirely to them, the first catalog and spectral atlas of known SySts by
\citet{Allen}, and the first simple geometrical modeling of their ionization
front \citep{Seaquist}.  Allen offered a clean classification criterium for
SySts: \textit{a binary star, combining a red giant (RG) and a companion hot
enough to sustain HeII (or higher ionization) emission lines.} The spectral
atlas by \citet{Munari02}, shows how the majority of SySts meeting this
criterium display in their spectra emission lines of at least the NeV, OVI
or FeVII ionization stages, requiring a minimum photo-ionization temperature
of 130,000 K \citep{Murset94}.  For sometime a main sequence star accreting
at a furious rate (10$^{-5}$$-$10$^{-4}$ M$_\odot$yr$^{-1}$) was advocated
as the hot component of several symbiotic stars \citep{Kenyon84, Joanna92},
but this scenario was later abandoned in favor of a WD.  Accretion onto a
main-sequence star must apply instead to {\em pre}-SySts systems like 17 Lep
\index{17 Lep} which are in the first phase of mass transfer \citep{Blind},
when the AGB progenitor of the future WD transfers mass to a main sequence
companion.  As indicated by satellite observations \citep[eg.][]{Munari94},
accretion alone cannot sustain the extreme luminosities
(10$^3$--10$^4$~L$_\odot$) encountered in most SySts \citep{Murset91}, and -
given its high efficiency - nuclear burning\index{nuclear burning} at the WD
surface must be invoked: burning 1 gr of hydrogen provides infact
6$\times$10$^{18}$ erg, while accreting the same 1 gr on a 1.3 M$_\odot$ WD
liberates 6$\times$10$^{17}$ erg, 6$\times$10$^{16}$ erg on a 0.5 M$_\odot$
WD, and only 2$\times$10$^{15}$ erg on a main-sequence star like the Sun. 

  \begin{figure}
    \includegraphics[scale=0.64]{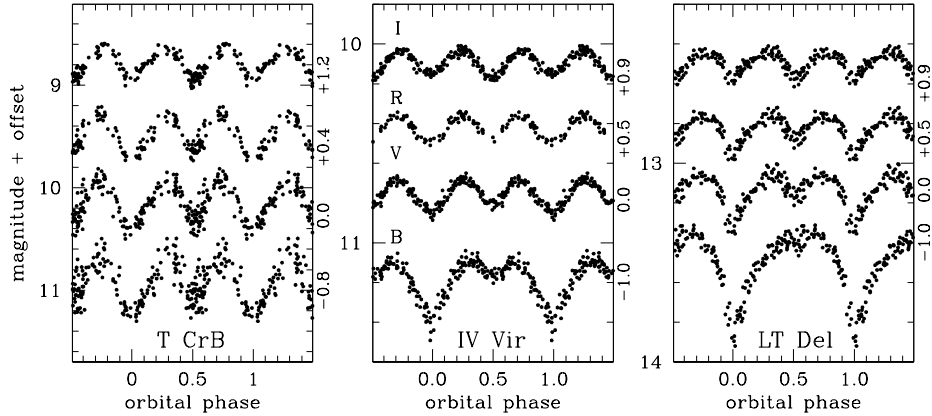}
    \caption[Roche lobe filling and irradiation in symbiotic stars]
        {BVRI lightcurves of three high-inclination symbiotic stars to
        illustrate the variable proportion between elipsoidal
        modulation\index{elipsoidal distortion} from red giant filling its
        Roche lobe (dominating the I-band flux; two maxima and two minima
        per orbital cycle) and irradiation\index{irradiation} of the same
        red giant by the hot WD companion (dominating the B band flux; one
        maximum and one minimum per orbital cycle).  \index{T Crb} \index{LT
        Del} \index{IV Vir}}
    \label{Munari2Fig:Fig2}
     \end{figure}

Homogeneity did not last for long however, and by the time of the next
catalog of SySts \citep{Belczynski}, systems with an excitation lower than
HeII, and evidence from X-rays for a neutron star (NS) rather than a WD,
begun percolating through the broadening classification criteria.  Currently
\citep{Mukai}, \textit{any binary where a WD or NS accretes enough material
from a RG companion such that this interaction can be detected at some
wavelength} is called a SySt.  The number of known SySts is rapidly
expanding \citep[$\sim$400,][]{Akras}, in particoular as a consequence of
surveys of the Galactic Bulge and Plane \citep{Corradi, Miszalski,
Rodriguez-Flores} and of galaxies in the Local Group \citep{Angeloni, Joanna14,
Joanna17}.

\section{Burning Symbiotic Stars}
\label{MunariSec2}

The amount of material burnt at the surface of the WD as been traditionally
considered equal to that continously accreted from the RG
\citep[eg.][]{Kenyon86}, so that neither the nuclear burning switches-off
(accretion too low), nor the envelope expands to red-giant dimension
(accretion too high).  The WD in burning SySts is however
radiating close to its Eddington limit\index{Eddington luminosity} and
accretion could fall quite shorter than required to replenish what
the WD burnt.  Either nuclear burning conditions are therefore met only
temporarely (followed by an accreting-only interval to refuel the envelope
of the WD) or the Eddington limit is circumvent by discrete accretion
episodes like that of a massive disk dumping mass onto the WD during
low-amplitude outbursts.  The burning WD emits profusely super-soft
X-rays\index{super-soft X-ray source}, and indeed several SySts have been
detected as SSS-sources, including C-1 in the Draco dwarf galaxy
\citep{Luna}\index{Draco C-1}.  Most however are not detected as
SSS-sources, the super-soft X-rays being absorbed locally by the abundant circumstellar
gas (a situation similar to the early evolution of novae, before the ejecta
dilute and turn optically thin to super-soft X-rays from the burning WD at
their center). The low impact made by accretion in burning SySts is confirmed by the
widespread absence of both flickering\index{flickering} and signature of
magnetic-driven accretion \citep{Sokoloski, Zamanov17}.

The SySts (mostly those of the burning type) are the only known class of
astronomical objects known to show OVI 1032, 1036 \AA\ Raman-scattered\index
{Raman scattering} by neutral hydrogen into a pair of broad emission
features at 6825, 7088 \AA\ \citep{Schmid89}.  A so far unique exception
seems to be the Sanduleak's Star\index{Sanduleak's star} in LMC, which
partnership with SySts has been questioned \citep{Angeloni}.  Other
Raman-scattered lines from HeII 940, 972 and 1025 \AA\, and NeVII 973 \AA\
have been proposed for SySts \citep{Lee}.  The simultaneous coexistence of
OVI and neutral hydrogen can be offered only by the very extended wind
(10$^2$ au) of a RG orbited within a few photospheric radii by a burning WD.

To accomodate a RG, the orbital separation in a SySt is measured in
astronomical units rather then solar radii of cataclysmic variables.  The
orbital periods range mostly from 1 to 4 yrs, with a maximum at 2/3 yrs and
an M5III spectral type for the RG.  Most burning SySts in our Galaxy seems
to belong to the metal-rich Bulge population \citep[Gaia will soon
tell]{Whitelock92} and are O-rich (M spectral types) as opposed to C-rich
(Carbon spectral types) in the Magellanic clouds, a fact related to the
lower metallicity of SMC/LMC and its impact on the amount of Carbon brought
up by the third dredge-up on the AGB.  Chemical abundances has been
recently derived for several of the brighest SySts from near-IR spectra. 
They should be treated with caution given the adopted over-simplifications
(thin, plane-parallel, static, LTE atmospheres) contrasting with the huge
complication of real RGB/AGB atmospheres (non-LTE, shocked, wind-supported,
macro-turbolent, and hugely 3D extended).  In about 15\% of known SySts, the
RG is a Mira, of an average M7III spectral type, and with a pulsation period
usually quite longer than for field Miras \citep{Whitelock03}.  To
accomodate the Mira well within its Roche lobe so to allow a regular
pulsation, the orbital period must be P$\geq$20 yrs.  At such a wide orbital
separation, no appreciable interaction would have occurred prior to the Mira
evolutionary stage, with the binary system classified as an isolated, field
RG.  The Miras in SySts frequently come with warm dust (D-type SySts, as
opposed to S-type with no detectable dust, and D'-type with cooler dust),
which is believed to be preferentially located in the shadow cone created by
the Mira itself, and possibly causing periodic obscurations along the
orbital motion (Whitelock 2003), even if an alternative location in the
collision zone between the winds from the Mira and the WD has been proposed
\citep{Hinkle}, at least for the Miras in symbiotic novae.

  \begin{figure}
    \includegraphics[scale=1.0]{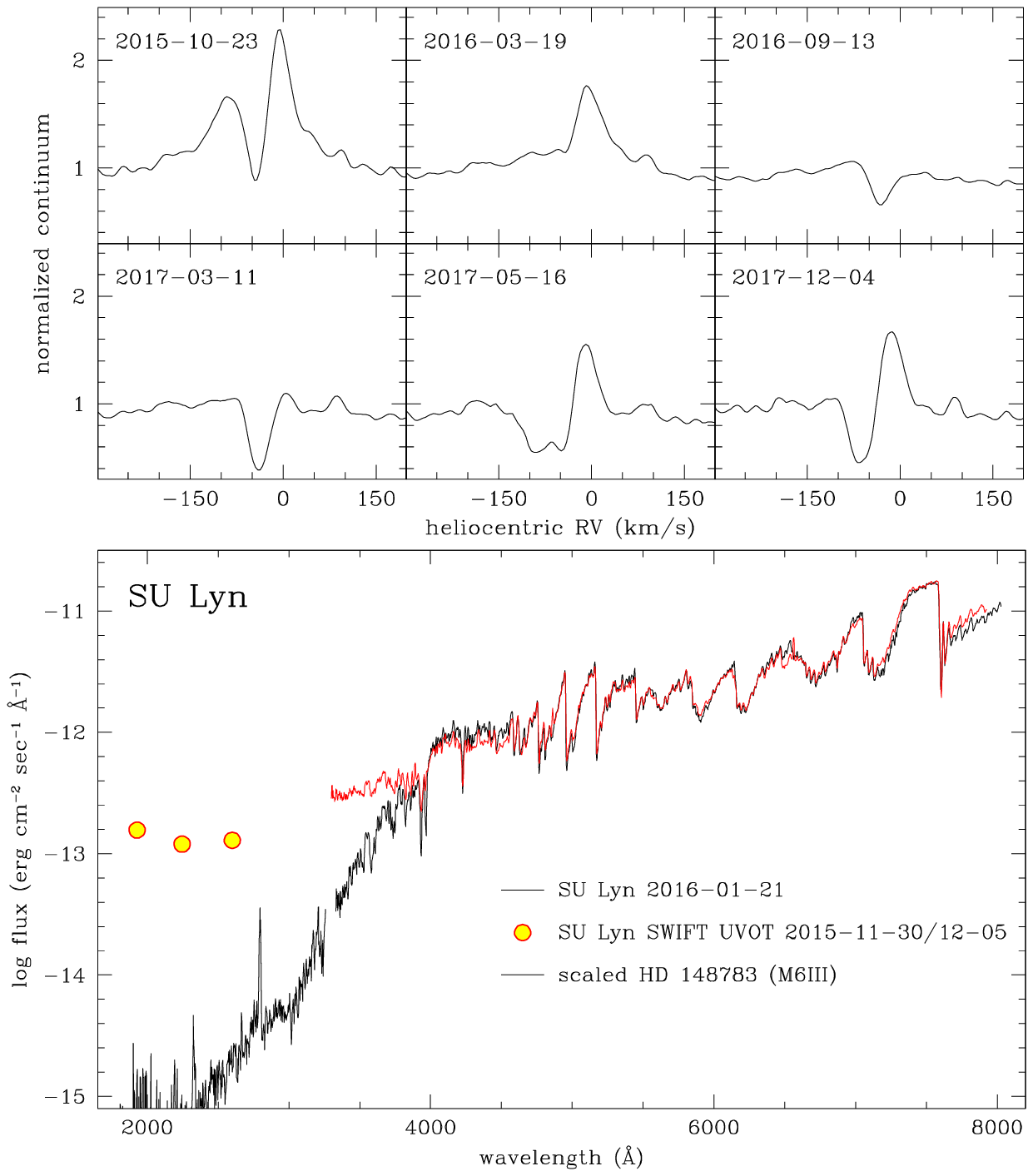}
    \caption[Signature of accretion in SU Lyn]
      {\textit{Bottom panel:} low resolution optical spectra of SU Lyn (red)
       reveals nothing different from those of field normal giant of the
       same spectral type (black).  The presence of an accreting WD is
       betrayed by the extra flux at far-blue ($\lambda$$\leq$3900 \AA) and
       UV wavelengths (red/yellow Swift UVOT dots).  \textit{Upper panel:}
       High resolution observations show feeble, structured, and highly
       variable emission in H$\alpha$ (and other lines as well), again
       disclosing the accreting nature of SU Lyn.}
    \label{Munari3Fig:Fig3}
     \end{figure}

The mass transfer from the RG to the WD of SySts can occur via either
capture from wind or Roche-lobe overflow\index{Roche-lobe overflow}.  SySts
with Roche-lobe filling RG present ellipsoidal distorted light-curves at
$I$-band or longer wavelengths, as illustrated in Figure~6.2, with an
amplitude depending on orbital inclination and spectral type.  For SySts with
accreting-only WD, the ellipsoidal modulation dominates all the way down to
bluest wavelengths (T CrB in Figure 6.2), while for burning SySts the
irradiation by the WD of the facing side of the RG is responsible for the
sinusoidal modulation dominating the bluest photometric bands.  The
amplitude of this modulation is proportional to the temperature and
luminosity of the burning WD (cf.  IV Vir and higher ionization LT Del in
Figure~6.2).  The strong orbital dependence of the emission in the Balmer
continuum (primarily responsible for the huge amplitude seen in the U band)
suggests that an important fraction of the irradiation effect resides in the
ionized gas between the WD and the RG \citep{Proga}, but at least in some
cases a direct increase in the surface temperature of the irradiated side of
the RG has been documented \citep{Chakrabarty, Munari16}.  There are SySts
showing no ellipsoidal distortion of the $I$ or $JHK$ light-curves while
presenting deep eclipses of the WD during outbursts (eg.  FG Ser).  Their RG
must resides well within the Roche lobe, and the WDs have therefore to
accrete from the wind, as it is the case for SySts containing a pulsating
Mira variable.  Hydrodynamic simulations \citep{Mohamed} show that the wind
can be confined within the RG's Roche lobe and strongly focused toward the
binary orbital plane.  Such a wind Roche-lobe overflow (WRLOF) can be so
efficient to allow the WD to accrete $\sim$50\% of the RG's mass loss and not
just the few \% typical of a Bondi-Hoyle-Littleton dynamical cross-section. 
The WRLOF, as other means of boosting the efficiency of mass accretion from
wind \citep{Bisikalo, Skopal15, Pan}, would also offer a way out to the
evolutionary paradox posed by the {\em yellow} SySts\index{yellow symbiotic
stars}.  They are a small group, including both burning and accreting-only
cases.  Their RG are G/K-type giants, with Halo kinematics and low
metallicity ([Fe/H]$\leq$$-1$), enriched in $s$-type elements\index{s-type
chemical element} (most notably barium\index{Barium star}), which are
normally brought to surface during third dredge-up\index{3rd dredge-up}
at the tip of the AGB.  They lack however the presence of unstable Tc
isotopes and are less luminous than the tip of the AGB, indicating and
extrinsic origin of the $s$-type elements, i.e.  pollution from the
progenitor of the current WD companion \citep{Jorissen, Pereira}.  Some of
these SySts are rotating at a significant fraction of their rotational
break-up velocities ($V_{\rm rot}\sin i$$\geq$100 km\,s$^{-1}$), suggesting a
massive transfer of both mass and angular momentum from the progenitor of
the current WD.  The distribution of orbital periods and eccentricities of
Barium SySts require that dynamically unstable mass transfer by Roche lobe
overflow (and the resulting common-envelope phase with its orbit shrinking
and circularization), is avoided and massive transfer of mass and angular
momentum be achieved via wind.

\section{Accreting-only Symbiotic Stars}
\label{MunariSec3}

All-sky surveys as well as pointed X-ray observations (with the {\it
Swift}\index{Swift X-ray satellite} satellite in particular) are
discovering a population of optically unconspicous RG that emits in hard
X-rays, a fact requiring them to pair in a binary system with a WD or a NS. 
Their relatively low X-ray ($\approx$0.1-10 L$_\odot$) and UV ($\approx$1-10
L$_\odot$) luminosities currently limits the serendipitous discovery to
systems within $\sim$1 kpc.  A WD companion is usually associated with a
luminosity larger in UV than in X-rays, while the reverse is true for a NS
(given its deepest potential well).  SySts are studied as potential
progenitors of type Ia supernovae, since the original proposal by
\citet{Munari92}, whose population synthesis was based on the estimated total
number of burning SySts in the Galaxy, i.e.  those easier to discover over
vast distances thanks to their spectacular emission line spectrum.  If
burning SyST are just the tip of an iceberg of momentarily quiet,
accreting-only SySts, the appeal of the symbiotic channel to SN Ia\index{SN
Ia} will be further boosted up.

  \begin{table}[Ht!]
    \begin{minipage}{300pt}
    \caption[Symbiotic Stars in X-rays]
      {Different types of X-ray emission observed in Symbiotic Stars, 
       their likely origin and some of the best known examples in each class.
       \index{AG Dra}
       \index{Draco C-1}
       \index{SMC-3}
       \index{Z And}
       \index{Mira AB}
       \index{AG Peg}
       \index{4 Dra}
       \index{T CrB}
       \index{NQ Gem}
       \index{CH Cyg}
       \index{MWC 560}
       \index{GX 1+4}
       \index{V934 Her}
       \index{4U 1954+31}}
    \label{MunariTable:Tab1}
    \addtolength\tabcolsep{2pt}
      \begin{tabular}{@{}c@{~~}l@{}c@{}}
        \hline \hline
        Type & Description & Examples\\        
        \hline
	$\alpha$ & Super-soft, photon energies $\leq$0.4 keV, & AG Dra, Draco C-1, SMC-3 \\
                 & hydrogen burning on WD surface & \\
        $\beta$  & Soft, photon energies $\leq$2.4 keV, & Z And, Mira AB, AG Peg \\
                 & colliding winds from WD and red giant \\
        $\delta$ & Hard, absorbed, with thermal emission& SU Lyn, 4 Dra, T CrB \\
                 & detectable at $\geq$2.4 keV, boundary layer\\
                 & between accretion disk and WD \\
        $\beta$/$\delta$ & characteristics of both $\beta$ and $\delta$ type & NQ Gem, CH Cyg\\
                 &simultaneously present, from colliding & MWC 560\\ 
                 & winds and disk/WD boundary layer \\
        $\gamma$ & Absorbed NS accretor, pulsed by NS & GX 1+4, V934 Her \\ 
                 & spin, optically thick Comptonized plasma& 4U 1954+31\\
        \hline \hline
      \end{tabular}
    \end{minipage}
  \end{table}

The subtle way these low-key, optically-quite and accreting-only SySts are
discovered is well epitomized by SU Lyn, a $V$$\sim$8 mag, M6III giant at
about 600 pc distance, completely unnoticed except for an old report about a
possible SRB variability (AGB stars with a poorly defined periodicity and
low amplitude).  Looking for optical counterparts of \textit{Swift}-BAT
sources, \citet{Mukai} noted that SU Lyn lied within the error box of one of
them.  Follow-up observations were organized with \textit{Swift} (to refine
the astrometric position of the BAT source and better characterize its UV
and X-ray emission) and with the Asiago spectrographs (to investigate if
SU Lyn optical spectra could betray peculiarities supporting a physical
association with the \textit{Swift}-BAT source).  Some results are
summarized in Figure~6.3.  While the optical spectrum of SU Lyn is identical
to that of a normal M6III giant, its bluest part ($\lambda$$\leq$3900 \AA)
shows a flux excess that extends to match the UV excess seen by
\textit{Swift} UVOT telescope and the soft and hard X-ray emission observed
by \textit{Swift} XRT and BAT instruments.  Only high-resolution Echelle
spectra can reveal a feeble, structured and quite variable emission in
H$\alpha$.

Similar tortuous paths affect the discovery of SySts hosting a NS, usually
named symbiotic X-ray binaries \citep[or SyXBs;][]{Masetti}.  SyXBs are
quite rare: among the $\sim$200 low-mass X-ray binaries (LMXBs) known in the
Galaxy, only $\sim$10 SyXBs cases are currently known.  Observationally,
these systems are characterized by appreciable X-ray emission
($\sim$10$^{32}$--10$^{34}$ erg s$^{-1}$) positionally associated with a RG
star which spectroscopically does not show any abnormal features, with the
possible exception of a continuum excess in the blue and ultraviolet ranges
(similar to what illustrated in Figure~6.3 for SU Lyn).  The X-ray emission is
pulsed (periods from 10$^2$ to 10$^4$ seconds, 4U 1954+319 being the slowest
at P$_{\rm spin}$$\sim$18,400 s), indicating that the NS is rotating slowly. 
The rotation period changes in response to accretion: for 4U
1954+319\index{4U 1954+31} \citet{Marcu} measured a strong spin-up of
$-$1.8$\times$10$^{-4}$ hr hr$^{-1}$ during outbursts and a spin-down of
2.1$\times$10$^{-5}$ hr hr$^{-1}$ in quiescence.  The level of X-ray
emission can vary up to four orders of magnitude, suggesting accretion from an
inhomogeneous stellar wind and possibly coupled with an highly elliptical
orbit of the accretor.  A notable outlier is GX 1+4\index{GX 1+4}, which
emits in X-rays up 10$^{37}$ erg s$^{-1}$, with P$_{\rm spin}$$\sim$120 s
(Chakrabarty and Roche 1997), and an optical spectrum (Munari and Zwitter
2002) quite similar to those of burning SySts.  A geometrically-thin and
optically-thick accretion disk heavily irradiated by the hard X-rays from
the central NS would provide the UV-source needed to photo-ionize the RG
wind: plasma diagnostic shows in fact that the gas in GX 1+4 is ionized by
thermal UV radiation ($T_{\rm ph}$$\sim$85,000 K) rather than the
non-thermal X-ray power law expected from the central accreting NS.

An handy classification of the main types of X-ray emission seen in SySts has
been introduced by \citet{Murset97} and expanded by \citet{Luna}
and \citet{Nunez}.  It is summarized in Table~1 with the name of
a few well known SySts in each class, and a compact description of the likely
origin for the X-rays.

\section{Different types of outburst in Symbiotic Stars}
\label{MunariSec4}

  \begin{figure}
    \includegraphics[scale=0.68]{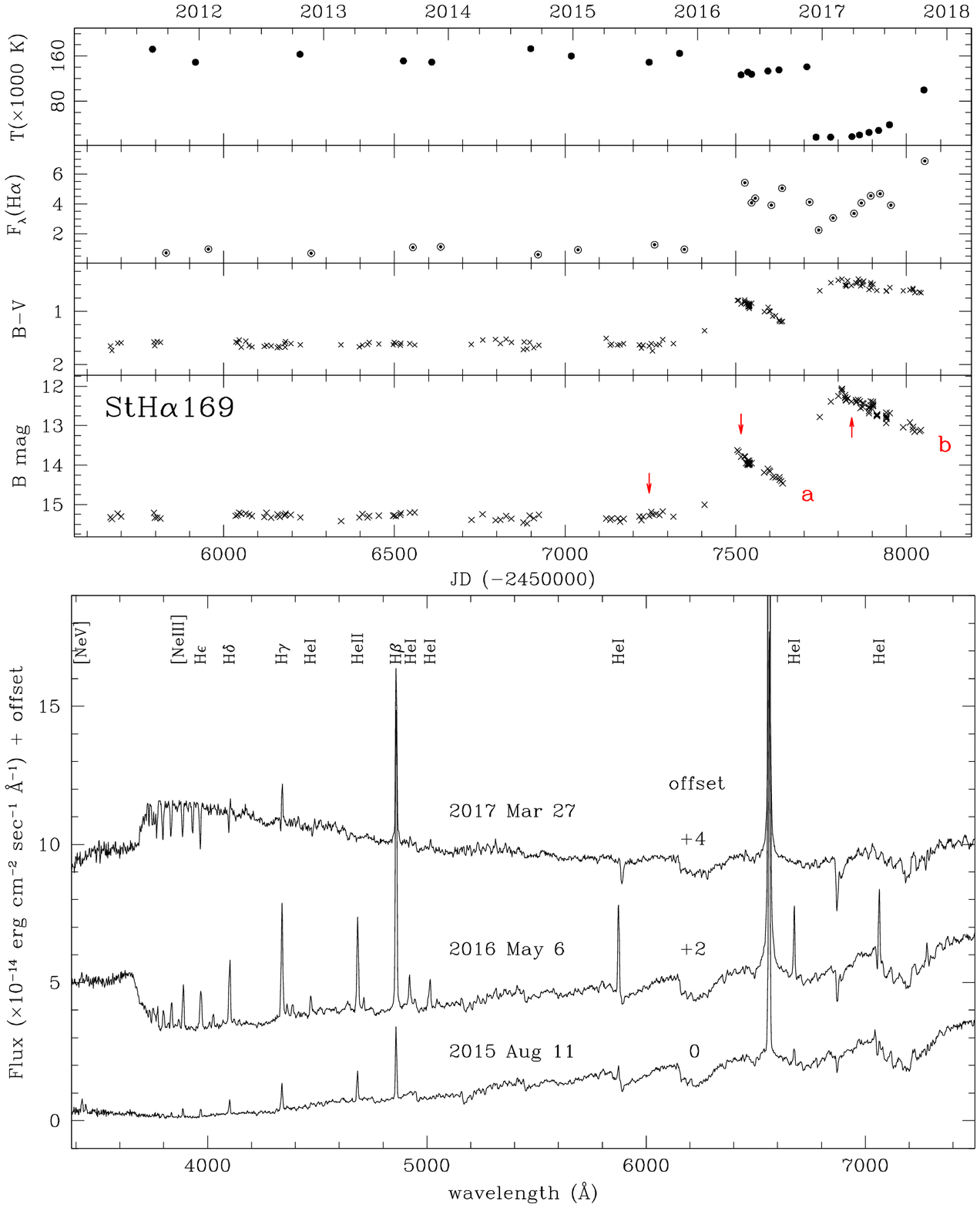}
    \caption[Different types of outbursts in StH$\alpha$ 169]
      {Two very different kinds of outburst in the same burning SySt: $a$ is
      an accretion event, $b$ an expansion of the burning shell once the
      accreted material reaches it.  The arrows points to times for the
      representative spectra in the bottom panel.  F$_\lambda$(H$\alpha$) is
      the integrated flux of H$\alpha$ (in units of 10$^{-12}$ erg cm$^{-2}$
      s$^{-1}$), and T the photo-ionization temperature of the hot source.}
    \label{Munari4Fig:Fig4}
     \end{figure}

\subsection{Z-And or Classical Type}.  

Normal SySts frequently enjoy outburts, that usual come in trains of a few
individual episodes separated by longer periods spent at quiescence.  The
example of CI Cyg\index{CI Cyg} in Figure~6.1 is indicative: three different
maxima ($\alpha$, $\beta$, $\gamma$), of declining strength and duration,
separated in time by $\sim$1 orbital period.  The previous train of multiple
maxima ended in 1979.  Also the spectral evolution depicted in Figure~6.1 is
quite typical: compared to quiescence, the ouburst spectrum is characterized
by a lower ionization, a stronger hot continuum veiling the RG molecular
spectrum, and a large increase in the integrated flux of the emission lines. 
The amplitude of the outburst is 2 or 3 mag in the $B$ band, and declines
toward the red.  The color evolution of CI Cyg in Figure~6.1 well illustrates
how the outburst status is barely detectable in the $I$-band, where the flux
is dominated by the RG at all phases.  This type of frequent and multiple
outbursts is called Z-type or Z-And type from Z Andromedae\index{Z And}, a
prototype SySt.  Jets (frequently seen bi-polar in high-res
spectra)\index{jet, bipolar} have been observed in about 12 symbiotic stars
(for 1/3 of them also spatially resolved; cf.  the spectacular images for R
Aqr by \citet{Schmid17}), and in most cases they are associated to Z-And
outbursts.  The projected jet velocities are of the order of 1000$-$1500
km\,s$^{-1}$, equivalent to the escape velocity from the region closest to the
central WD, with the noteworthy exception of MWC~560\index{MWC 560} where
velocities $V_{\rm ej}$$\geq$6,000 km s$^{-1}$ were observed
\citep{Tomov90}.  In response to the great differences seen in the jets from
one object to another, a variety of launching mechanisms have been proposed
and modelled \citep[eg.][]{Stute, Skopal09, Tomov11}.  Whereas RGs in SySts
rotate faster than field RGs and appear synchronized\index{rotation
synchronization} with orbital period (P$_{\rm rot}$ $\simeq$ P$_{\rm orb}$),
those in systems emitting jets seem to rotate faster at P$_{\rm
rot}$$<$P$_{\rm orb}$ \citep{Zamanov12}.

As of the causes of Z-And type outburst, a great variety of different
mechanisms have been invoked \citep[eg.][]{Bisikalo, Tomov11, Skopal11,
Ramsay, deValBorro} like a sudden increase in the mass-transfer rate from
the RG, either triggered by intrinsic variability of the RG or its passage
at periastron; the formation of an optically thick, cool, disk-shaped zone
around the WD equator as a consequence of enhanced wind from the WD; an
enhanced wind from the WD which leads instead to the disruption of the inner
part of the accretion disk with the formation of hollow cones around the WD
axis of rotation and thus to the appearance of collimated outflows; changes
in the kinematical regime of colliding winds from the WD and the RG, etc.  A
coordinated effort between X-ray, ultraviolet, optical, and radio
observations to follow in detail and over all the relevant phases the Z-And
outbursts of at least a few SySts seems required to rise firm constraints
useful in guiding future modeling efforts.

Basically, the explanations for Z-And type outbursts tend to cluster into
two broad categories: ($a$) release of potential energy from extra-accreted
matter, or ($b$) shift to longer wavelengths of the emission from WD
burning shell of the WD, expanding in radius as response to an increase in the mass
accretion rate.  Both modes could occur in succession in the same object, as
illustrated in Figure~6.4 by the 2016/17 outburst of StH$\alpha$
169\index{StHa 169}.

  \begin{figure}
    \includegraphics[scale=0.69]{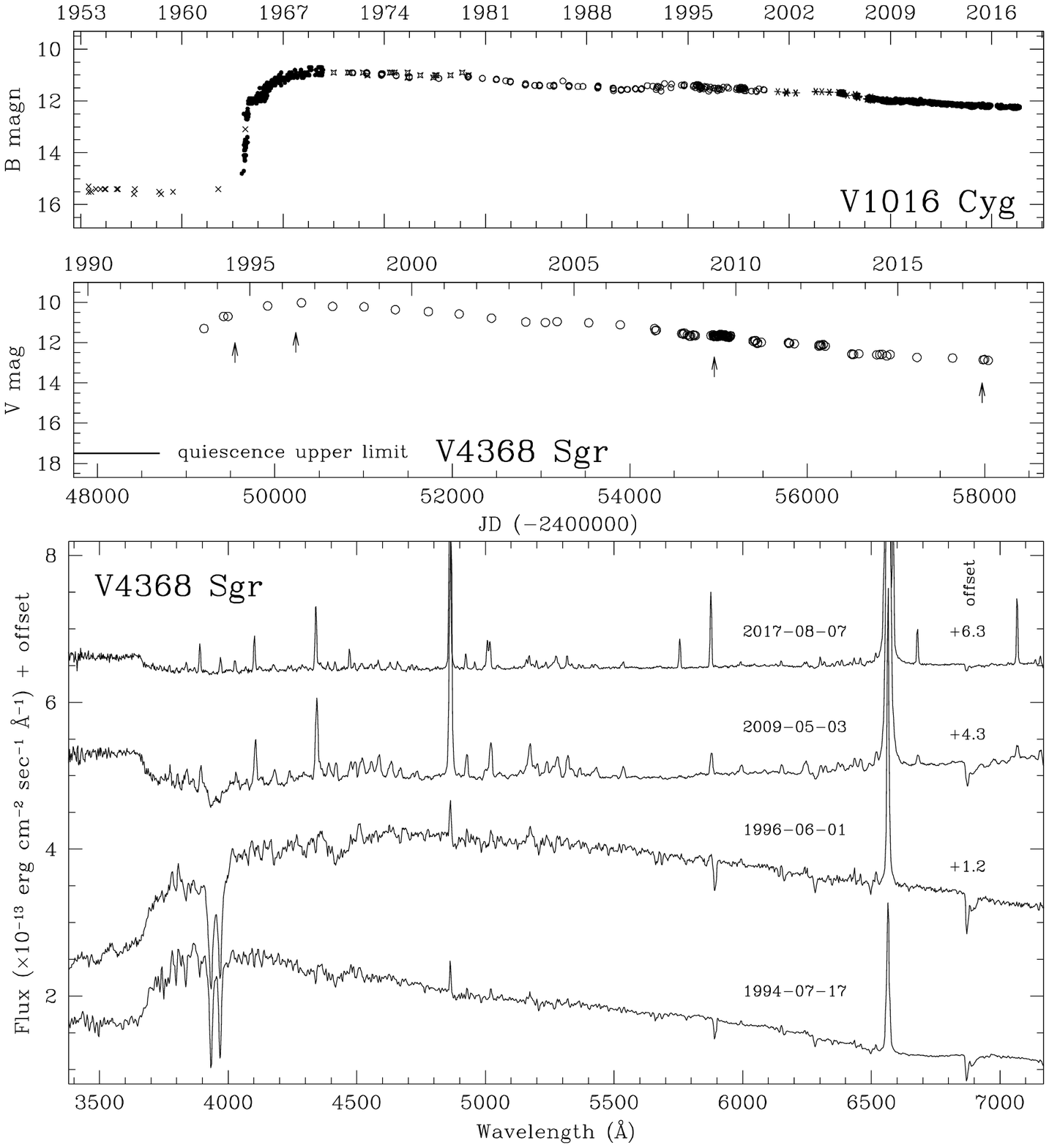}
    \caption[Photometric and spectroscopic evolution of symbiotic novae]
      {The $V$-band light-curve of the symbiotic nova V4368 Sgr (= Wakuda's
      object) is given in the middle panel, with marked by arrows the epochs
      of the sample spectra shown in the bottom panel.  For comparison, the
      upper panel displays the light-curve of V1016 Cyg, another SyN.}
    \label{Munari5Fig:Fig5}
     \end{figure}

\subsection{Symbiotic Novae}.  

A very few ($\sim$10) Symbiotic Novae (SyN) have been seen to erupt in
historical times in our Galaxy, while others were already in outburst when
discovered as SySts.  They should not be confused with the novae erupting in
symbiotic binaries described below.  The outbursts of SyN last about a
century, are of large amplitude and only one eruption has been recorded,
with the possible exception of BF Cyg that shortly after returning to
quiescence from the outburst initiated in 1894 \citep{Leibowitz}, it started
a new SyN cycle in 2006 and currently is still at maximum brightness.

The typical photometric and spectroscopic evolution of SyN are illustrated
in Figure~6.5.  A large rise in brightness takes the star, in about $\sim$1
year, from faint anonymity to bright spotlight, from where an extremely slow
decline needs about a century to return the system to quiescence level.  The
spectral evolution is equally slow.  During the rise toward maximum, the
spectrum cools up to an F-type supergiant with feeble emission lines limited
to Balmer and FeII.  During the decline, the F-type continuum weakens, a
nebular continuum takes over and the emission lines grow in intensity and
ionization degree.  In few cases (eg.  HM Sge\index{HM Sge} and V1016
Cyg\index{V1016 Cyg}), the phase dominated by the F-type supergiant
continuum is probably too short and, when transiting at maximum brightness,
the spectrum of the SyN is already dominated by a nebular continuum and
strong emission lines.

A SyN outburst could even be the event that initiate a symbiotic-cycle in
the life of a RG+WD binary, coming after a long and quiet period of
accretion only.  Under the high mass-transfer rates allowed by WRLOF or by
plain Roche-lobe filling, the material should pile up in non-degenerate form
on the surface of a nonmassive and hot WD.  Upon reaching conditions for hydrogen
burning, this will proceed under thermal equilibrium avoiding the TNR of a
classical nova and the consequent violent mass ejection.  To adjust to the
large nuclear luminosity produced at its base, the nondegenerate envelope
expands to supergiant dimensions.  The absence of massive ejection retains
most of the mass in the WD envelope, which keeps burning under stable
conditions for a long time, in excess of the $\sim$century a SyN takes to
return to quiescence.  AG Peg\index{AG Peg} has only recently returned to
pre-SyN brightness after the SyN outburst initiated around 1850: now it is a
normal-burning SySt that experiences normal Z-And type outbursts
\citep{Tomov16, Ramsay}.  If accretion cannot keep pace with hydrogen
depletion by nuclear burning, sooner or later the shell on the WD in AG Peg
will slim under the critical value, the burning will stop, and the star will
move back to the anonymity typical of accreting-only SySts.  After quietly
accreting for an appropriately long interval, it will be ready for the next
symbiotic cycle to be initiated by a new century-long SyN outburst.

  \begin{figure}
    \includegraphics[scale=0.60]{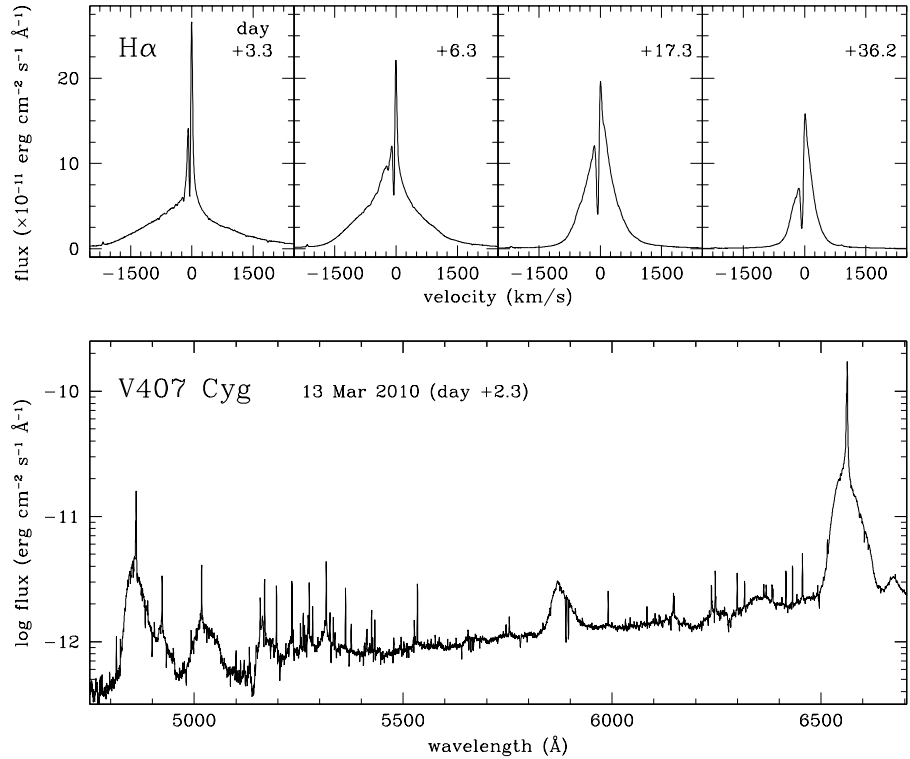}
    \caption[Flash-ionized wind and deceleration of nova ejecta in V407 Cyg]
        {The peculiar spectral evolution of a nova erupting within a
         symbiotic binary (NwSySt) is well illustrated by the 2010 outburst
         of V407\index{V407 Cyg} Cyg.  \textit{Bottom:} portion of an
         Echelle spectrum obtained +2.3 days past optical maximum, with the
         simultaneous presence of very sharp (FWHM$\leq$20 km s$^{-1}$) and
         broad lines (FWHM$\sim$2700 km s$^{-1}$).  Sharp lines are produced
         by recombination within the flash-ionized wind of the red giant. 
         Broad lines comes from the expanding ejecta of the nova. 
         \textit{Top:} sample of H$\alpha$ profiles at various epochs
         showing the quick disappearance of the narrow component, the
         violent deceleration of the broad one, and the persistent presence
         of a narrow absorption originating in the outer wind of the RG
         unperturbed by the nova outburst.}
    \label{Munari6Fig:Fig6}
     \end{figure}

\subsection{Novae Erupting Within Symbiotic Stars}.  

Under proper balance between mass loss and gain, the WDs of symbiotic stars
can grow in mass toward the Chandrashekar limit\index{Chandrashekar limit},
with the bright prospect of concluding their life with a spectacular
explosion as type Ia supernovae \citep{Munari92}.  Approaching that limit,
the WDs become so massive that normal nova explosions could occur on such a
short time scale that more than one has been observed in historical times. 
Most famous recurrent novae\index{recurrent nova} among SySts are RS
Oph\index{RS Oph} (7 outbursts), V745 Sco\index{V745 Sco} (3), T CrB\index{T
CrB} (2), and V3890 Sgr\index{V3890 Sgr} (also 2 outbursts).

A nova erupting within a SySt (NwSySt) evolves quite differently from
a classical one.  When the TNR culminates with an
intense UV flash (Starrfield et al., 2016), the RG wind absorbs most/all of
it, get ionized, and soon start glowing under recombination.  The NwSySt is
taken almost instantaneously to peak optical brightness, whereas in
classical novae the UV flash disperses in the surrounding emptiness and goes
unnoticed hours/days before the nova is discovered, and well before peak
brightness is attained at the time of maximum expansion for the
pseudo-photosphere of the optically thick ejecta.  Given the large electron
density in the wind of the RG at the distance it is orbited by the WD
(10$^6$--10$^8$ cm$^{-3}$), the recombination from the UV flash proceeds
rapidly in NwSySts ($e$-folding time 3--6 days).  The recombining wind is not
kinematically perturbed, consequently the lines it emits remain very sharp
(FWHM $\simeq$20 km s$^{-1}$).  In the meantime, material is ejected at high
velocity from the central nova (FWHM of thousands km s$^{-1}$), producing
very wide emission lines (of the He/N nova type) co-existing with the narrow ones
(Figure~6.6, bottom panel).  The fast ejecta ram onto the pre-existing wind of
the RG, and upon sweeping it up they are violently decelerated causing a
rapid narrowing of the broad-lines profiles ($e$-folding time of a few days;
Figure~6.6 top panel).  $\gamma$-rays in the GeV range are then produced as a
consequence of the violent shock.  The best documented NwSySt eruption is
probably that of 2010 for the symbiotic Mira V407 Cyg\index{V407 Cyg}
\citep{Munari11, Pan}.

  \bibliography{percolation}\label{refs}
  \bibliographystyle{cambridgeauthordate}
  
 \copyrightline{} 
 \printindex
    
\end{document}